\documentclass{sig-alternate}

\newcommand{\barra}[1]{\overline{#1}}

\newdef{definition}{Definition}
\newdef{theorem}{Theorem}

\begin{document}



\pagenumbering{arabic}
\setpagenumber{79}

\title{Synthesis of ``correct'' adaptors for protocol enhancement in component-based systems\titlenote{This work is an extended and
revisited version of~\cite{wcat04}.}}

\numberofauthors{2}

\author{
\alignauthor Marco Autili, Paola Inverardi,\\ Massimo Tivoli\\
       \affaddr{University of L'Aquila}\\
       \affaddr{Dip. Informatica}\\
       \affaddr{via Vetoio 1, 67100 L'Aquila}\\
       \email{\{marco.autili, inverard, tivoli\}@di.univaq.it}
\alignauthor David Garlan\\
       \affaddr{Carnegie Mellon University}\\
       \affaddr{5000 Forbes Avenue}\\
       \affaddr{Pittsburgh, PA 15213-3891}\\
       \email{garlan@cs.cmu.edu}
}

\maketitle

\begin{abstract}
Adaptation of software components is an important issue in
\emph{Component Based Software Engineering} (CBSE). Building a
system from reusable or \emph{Commercial-Off-The-Shelf} (COTS)
components introduces a set of problems, mainly related to
compatibility and communication aspects. On one hand, components may
have incompatible interaction behavior. This might require to
restrict the system's behavior to a subset of safe behaviors. On the
other hand, it might be necessary to enhance the current
communication protocol. This might require to augment the system's
behavior to introduce more sophisticated interactions among
components. We address these problems by enhancing our architectural
approach which allows for detection and recovery of incompatible
interactions by synthesizing a suitable coordinator. Taking into
account the specification of the system to be assembled and the
specification of the protocol enhancements, our tool (called
\emph{SYNTHESIS}) automatically derives, in a compositional way, the
glue code for the set of components. The synthesized glue code
implements a software coordinator which avoids incompatible
interactions and provides a protocol-enhanced version of the
composed system. By using an assume-guarantee technique, we are able
to check, in a compositional way, if the protocol enhancement is
consistent with respect to the restrictions applied to assure the
specified safe behaviors.
\end{abstract}

\section{Introduction} \label{introduction}

Adaptation of software components is an important issue in
\emph{Component Based Software Engineering} (CBSE). Nowadays, a
growing number of systems are built as composition of reusable or
\emph{Commercial-Off-The-Shelf} (COTS) components. Building a system
from reusable or from COTS~\cite{Szy98} components introduces a set
of problems, mainly related to communication and compatibility
aspects. Often, components may have incompatible interaction
behavior. This might require to restrict the system's behavior to a
subset of safe behaviors. For example, restrict to the subset of
deadlock-free behaviors or, in general, to a specified subset of
desired behaviors. Moreover, it might be necessary to enhance the
current communication protocol. This requires augmenting the
system's behavior to introduce more sophisticated interactions among
components. These enhancements (i.e.: protocol transformations)
might be needed to achieve dependability, to add extra-functionality
and/or to properly deal with system's architecture updates (i.e.:
components aggregating, inserting, replacing and removing).

We address these problems enhancing our architectural approach which
allows for detection and recovery of incompatible interactions by
synthesizing a suitable coordinator~\cite{bertinoro,cbse7,savcbs03}.
This coordinator represents a initial glue code. So far, as reported
in~\cite{bertinoro,cbse7,savcbs03}, the approach only focussed on
the restriction of the system's behavior to a subset of safe (i.e.:
desired) behaviors. In this paper, we propose an extension that
makes the coordinator synthesis approach also able to automatically
transform the coordinator's protocol by enhancing the initial glue
code. We implemented the whole approach in our \emph{SYNTHESIS}
tool~\cite{cbse7,companion}\\
(\emph{http://www.di.univaq.it/tivoli/SYNTHESIS/synthesis.html}).
In~\cite{companion}, which is a companion paper, we also apply
\emph{SYNTHESIS} to a real-scale context. Since in this paper we are
focusing only on the formalization of the approach, in
Section~\ref{methodformalization}, we will simply refer to an
explanatory example and we will omit implementation details which
are completely described in~\cite{companion}.

Starting from the specification of the system to be assembled and
from the specification of the desired behaviors, \emph{SYNTHESIS}
automatically derives the initial glue code for the set of
components. This initial glue code is implemented as a coordinator
mediating the interaction among components by enforcing each desired
behavior as reported in~\cite{bertinoro,cbse7,savcbs03}.
Subsequently, taking into account the specification of the needed
protocol enhancements and performing the extension we formalize in
this paper, \emph{SYNTHESIS} automatically derives, in a
compositional way, the enhanced glue code for the set of components.
This last step represents the contribution of this paper with
respect to~\cite{bertinoro,cbse7,savcbs03}. The enhanced glue code
implements a software coordinator which avoids not only incompatible
interactions but also provides a protocol-enhanced version of the
composed system. More precisely, this enhanced coordinator is the
composition of a set of new coordinators and components assembled
with the initial coordinator in order to enhance its protocol. Each
new component represents a wrapper component. A wrapper intercepts
the interactions corresponding to the initial coordinator's protocol
in order to apply the specified enhancements without
modifying~\footnote{This is needed to achieve compose-ability in
both specifying the enhancements and implementing them.} the initial
coordinator and the components in the system. The new coordinators
are needed to assemble the wrappers with the initial coordinator and
the rest of the components forming the composed system. It is
worthwhile noticing that, in this way, we are readily compose-able;
we can treat the enhanced coordinator as a new \emph{composite}
initial coordinator and enforce new desired behaviors as well as
apply new enhancements. This allows us to perform a protocol
transformation as composition of other protocol transformations by
improving on the reusability of the synthesized glue code.

When we apply the specified protocol enhancements to produce the
enhanced coordinator, we might re-introduce incompatible
interactions avoided by the initial coordinator. That is, the
enhancements do not hold the desired behaviors specified to produce
the initial coordinator. In this paper, we also show how to check if
the protocol enhancement holds the desired behaviors enforced
through the initial coordinator. This is done, in a compositional
way, by using an assume-guarantee technique~\cite{Clarke99}.

The paper is organized as follows: Section~\ref{related} discusses
related work. Section~\ref{background} introduces background notions
helpful to understand our approach. Section~\ref{methoddescription}
illustrates the technique concerning the enhanced coordinator
synthesis. Section~\ref{methodformalization} formalizes the
coordinator synthesis approach for protocol enhancement in
component-based systems, and uses a simple explanatory example to
illustrate the ideas. Section~\ref{check} formalizes the technique
used to check the consistency of the applied enhancements with
respect to the enforced desired behaviors. Section~\ref{conclusion}
discusses future work and concludes.

\section{Related work} \label{related}

The approach presented in this paper is related to a number of other
approaches that have been considered in the literature. The most
closely related work is the scheduler synthesis for discrete event
physical systems using supervisory control~\cite{supervisory1}. In
those approaches system's allowable executions are specified as a
set of traces. The role of the supervisory controller is to interact
with the running system in order to cause it to conform to the
system specification. This is achieved by restricting behavior so
that it is contained within the desired behavior. To do this, the
system under control is constrained to perform events only in strict
synchronization with a synthesized \emph{supervisor}. The synthesis
of a supervisor that restrict behaviors resembles one aspect of our
approach defined in Section~\ref{methoddescription}, since we also
eliminate certain incompatible behaviors through synchronized
coordination. However, our approach goes well beyond simple
behavioral restriction, also allowing augmented interactions through
protocol enhancements.

Recently a reasoning framework that supports modular checking of
behavioral properties has been proposed for the compositional
analysis of component-based
design~\cite{interface_automata,converter_synthesis}.
In~\cite{interface_automata}, they use an automata-based approach to
capture both input assumptions about the order in which the methods
of a component are called, and output guarantees about the order in
which the component calls external methods. The formalism supports
automatic compatibility checks between interface models, where two
components are considered to have compatible interfaces if there
exists a legal environment that lets them correctly interact. Each
legal environment is an adaptor for the two components. However,
they provide only a consistency check among component interfaces,
but differently from our work do not treat automatic synthesis of
\emph{adaptors} of component interfaces.
In~\cite{converter_synthesis}, they use a game theoretic approach
for checking whether incompatible component interfaces can be made
compatible by inserting a converter between them which satisfies
specified requirements. This approach is able to automatically
synthesize the converter. The idea they develop is the same idea we
developed in our precedent works~\cite{bertinoro,cbse7,savcbs03}.
That is the restriction of the system's behavior to a subset of safe
behaviors. Unlike the work presented in this paper, they are only
able to restrict the system's behavior to a subset of desired
behaviors and they are not able to augment the system's behavior to
introduce more sophisticated interactions among components.

Our research is also related to work in the area of protocol adaptor
synthesis~\cite{Yel97}. The main idea of this approach is to modify
the interaction mechanisms that are used to glue components together
so that compatibility is achieved. This is done by integrating the
interaction protocol into components. However, they are limited to
only consider syntactic incompatibilities between the interfaces of
components and they do not allow the kind of protocol
transformations that our synthesis approach supports.

In other previous work, of one of the authors, we showed how to use
formalized protocol transformations to augment connector
behavior~\cite{wrappers_formalization}. The key result was the
formalization of a useful set of connector protocol enhancements.
Each enhancement is obtained by composing wrappers. This approach
characterizes wrappers as modular protocol transformations. The
basic idea is to use wrappers to enhance the current connector
communication protocol by introducing more sophisticated
interactions among components. Informally, a wrapper is new code
that is interposed between component interfaces and communication
mechanisms. The goal is to alter the behavior of a component with
respect to the other components in the system, without actually
modifying the component or the infrastructure itself. While this
approach deals with the problem of enhancing component interactions,
unlike this work it does not provide automatic support for composing
wrappers, or for automatically eliminating incompatible interaction
behaviors.

In other previous work, by two of the authors, we showed how to
apply protocol enhancements by dealing with components that might
have syntactic incompatibility of interfaces~\cite{CMU_techrep}.
However the approach described in~\cite{CMU_techrep} is limited only
to consider deadlock-free coordinator.

\section{Background} \label{background}

In this section we discuss the background needed to understand the
approach that we formalize in Section~\ref{methoddescription}.

\subsection{The reference architectural style} \label{CBA_architecture}

The starting point for our work is the use of a formal
architectural model of the system representing the components to
be integrated and the connectors over which the components will
communicate~\cite{GS96}. To simplify matters we will consider the
special case of a generic layered architecture in which components
can request services of components below them, and notify
components above them. Specifically, we assume each component has
a top and bottom interface. The top (bottom) interface of a
component is a set of top (bottom) ports. Connectors between
components are synchronous communication channels defining top and
bottom ports.

Components communicate by passing two types of messages:
notifications and requests. A notification is sent downward, while
a request is sent upward. We will also distinguish between two
kinds of components (i) \emph{functional components} and (ii)
\emph{coordinators}. Functional components implement the system's
functionality, and are the primary computational constituents of a
system (typically implemented as COTS components). Coordinators,
on the other hand, simply route messages and each input they
receive is strictly followed by a corresponding output. We make
this distinction in order to clearly separate components that are
responsible for the functional behavior of a system and components
that are introduced to aid the integration/communi\-cation
behavior.

Within this architectural style, we will refer to a system as a
\emph{Coordinator-Free Architecture} (CFA) if it is defined
without any coordinators. Conversely, a system in which
coordinators appear is termed a \emph{Coordinator-Based
Architecture} (CBA) and is defined as \emph{a set of functional
components directly connected to one or more coordinators, through
connectors, in a synchronous way}.

\begin{figure}[h]
\centering \epsfig{file=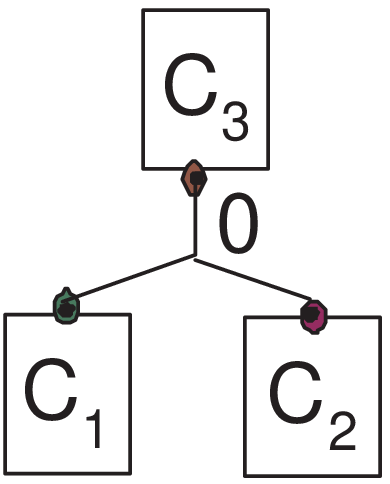, height=1.5cm,
width=1.5cm} \epsfig{file=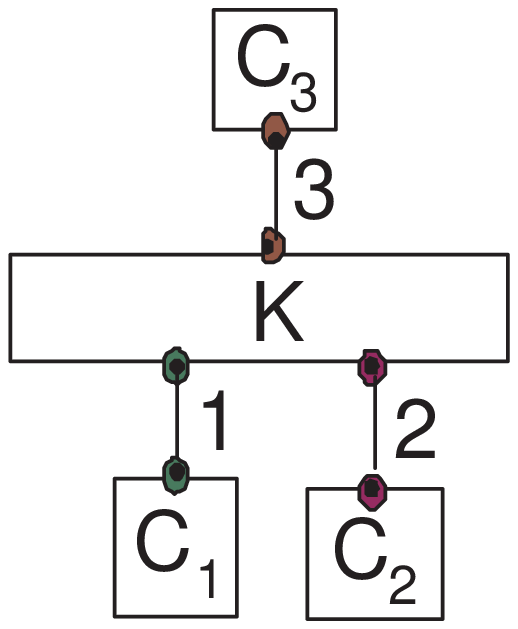, height=1.5cm,
width=2cm} \caption{A sample of a CFA and the corresponding CBA}
\label{cba_arch_sample}
\end{figure}

\noindent Figure~\ref{cba_arch_sample} illustrates a CFA (left-hand
side) and its corresponding CBA (right-hand side). $C_1$, $C_2$ and
$C_3$ are functional components; $K$ is a coordinator. The
communication channels identified by $0$, $1$, $2$ and $3$ are
connectors.

\subsection{Configuration formalization} \label{configuration_formalization}

To formalize the behavior of a system we use \emph{High level
Message Sequence Charts} (HMSCs) and \emph{basic Message Sequence
Charts} (bMSCs)~\cite{itu:msc} specification of the composed system.
From it, we can derive the corresponding CCS (\emph{Calculus of
Communicating Systems}) processes~\cite{milner:ccs} (and hence
\emph{Labeled Transitions Systems} (LTSs)) by applying a suitable
adaptation of the translation algorithm presented
in~\cite{implied_scenarios}. HMSC and bMSC specifications are useful
as input language, since they are commonly used in software
development practice. Thus, CCS can be regarded as an internal
specification language. Later we will see an example of derivation
of LTSs from a bMSCs and HMSCs specification
(Section~\ref{first_step}).

To define the behavior of a \emph{composition} of components, we
simply place in parallel the LTS descriptions of those components,
hiding the actions to force synchronization. This gives a CFA for
a set of components.

We can also produce a corresponding CBA for these components with
equivalent behavior by automatically deriving and interposing a
"\emph{no-op}" coordinator between communicating components. That
coordinator does nothing (at this point), it simply passes events
between communicating components (as we will see later the
coordinator will play a key role in restricting and augmenting the
system's interaction behavior). The "\emph{no-op}" coordinator is
automatically derived by performing the algorithm described
in~\cite{bertinoro,cbse7,savcbs03}.

Formally, using CCS we define the CFA and the CBA for a set of
components $C_1,..,C_n$ as follows:

\begin{definition} \label{cfaccs_def} \emph{Coordinator
Free Architecture (CFA)}
\\ $CFA\equiv(C_1 \mid C_2 \mid ... \mid C_n) \backslash
\bigcup_{i=1}^{n} Act_{C_i}$ where for all\\ $i=1,..,n$, $Act_{C_i}$
is the action set of the CCS process $C_i$.
\end{definition}

\begin{definition} \label{cbaccs_def} \emph{Coordinator
Based Architecture (CBA)}
\\ $CBA\equiv(C_1[f^0_1] \mid C_2[f^0_2] \mid ... \mid C_n[f^0_n]
\mid K) \backslash \bigcup_{i=1}^{n} Act_{C_i}[f^0_i]$ where for all
$i=1,..,n$, $Act_{C_i}$ is the action set of the CCS process $C_i$,
and $f^0_i$ are relabelling functions such that
$f^0_i(\alpha)=\alpha[i/0]$ for all $\alpha\in~Act_{C_i}$; $K$ is
the CSS process corresponding to the automatically synthesized
coordinator.
\end{definition}

\noindent By referring to~\cite{milner:ccs}, $\mid$ is the
\emph{parallel composition} operator and $\backslash$ is the
\emph{restriction} operator. There is a finite set of visible
actions $Act = \{a_i,\barra{a}_j,b_h,\barra{b}_k,...\}$ over which
$\alpha$ ranges. We denote by $\barra{\alpha}$ the action
complement: if $\alpha=a_j$, then $\barra{\alpha}=\barra{a}_j$,
while if $\alpha=\barra{a}_j$, then $\barra{\alpha}=a_j$. By
$\alpha[i/j]$ we denote a substitution of $i$ for $j$ in $\alpha$.
If $\alpha=a_j$, then $\alpha[i/j]=a_i$. Each $Act_{C_i}\subseteq
Act$. By referring to Figure~\ref{cba_arch_sample}, 0 identifies the
only connector (i.e: communication channel) present in the CFA
version of the composed system. Each relabelling function $f^0_i$ is
needed to ensure that the components $C_1,..C_n$ no longer
synchronize directly. In fact by applying these relabelling
functions (i.e.: $f^0_i$ for all $i$) each component $C_i$
synchronizes only with the coordinator $K$ through the connector $i$
(see right-hand side of Figure~\ref{cba_arch_sample}).

\subsection{Automatic synthesis of failure-free\\ coordinators} \label{old_approach}

In this section, we simply recall that from the MSCs specification
of the CFA and from a specification of desired behaviors, the old
version of \emph{SYNTHESIS} automatically derives the corresponding
deadlock-free CBA which satisfies each desired behavior. This is
done by synthesizing a suitable coordinator that we call
failure-free coordinator. Informally, first we synthesize a "no-op"
coordinator. Second, we restrict its behavior by avoiding possible
deadlocks and enforcing the desired behaviors. Each desired behavior
is specified as a \emph{Linear-time Temporal Logic} (LTL) formula
(and hence as the corresponding \emph{B\"{u}chi
Automaton})~\cite{Clarke99}. Refer
to~\cite{bertinoro,cbse7,savcbs03} for a formal description of the
old approach and for a brief overview on the old version of our
\emph{SYNTHESIS} tool.

\section{Method description} \label{methoddescription}

In this section, we informally describe the extension of the old
coordinator synthesis approach~\cite{bertinoro,cbse7,savcbs03} that
we formalize in Section~\ref{methodformalization} and we implemented
in the new version of the \emph{SYNTHESIS} tool. The extension
starts with a deadlock-free CBA which satisfies specified desired
behaviors and produces the corresponding protocol-enhanced CBA.

The problem we want to face can be informally phrased as follows:
\emph{let $P$ be a set of desired behaviors, given a deadlock-free
and $P$-satisfying CBA system $S$ for a set of black-box components
interacting through a coordinator $K$, and a set of coordinator
protocol enhancements $E$, if it is possible, automatically derive
the corresponding enhanced, deadlock-free and $P$-satisfying CBA
system $S^{\prime}$}.

\begin{figure}
\centering \epsfig{file=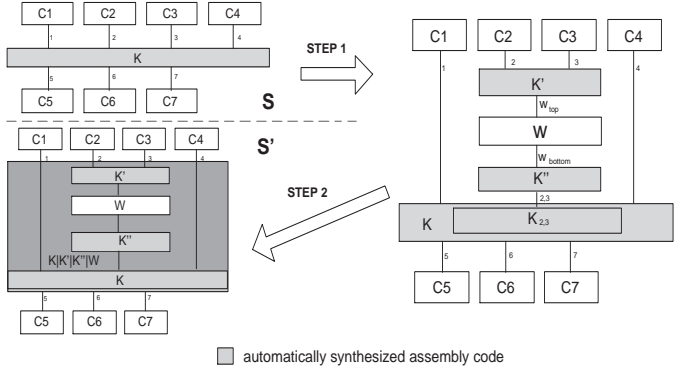, height=5cm, width=9cm}
\caption{2 step method} \label{fig:figura1}
\end{figure}

We are assuming a specification of: i) $S$ in terms of a description
of components and a coordinator as LTSs, ii) $P$ in terms of a set
of B\"{u}chi Automata, and of iii) $E$ in form of bMSCs and HMSCs
specification. In the following, we discuss our method proceeding in
two steps as illustrated in Figure~\ref{fig:figura1}.

In the first step, by starting from the specification of $P$ and
$S$, if it is possible, we apply each protocol enhancement in $E$.
This is done by inserting a wrapper component $W$ between $K$ (see
Figure~\ref{fig:figura1}) and the portion of $S$ concerned with the
specified protocol enhancements (i.e.: the set of $C2$ and $C3$
components of Figure~\ref{fig:figura1}). It is worthwhile noticing
that we do not need to consider the entire model of $K$ but we just
consider the \emph{"sub-coordinator"} which represents the portion
of $K$ that communicates with $C2$ and $C3$ (i.e.: the
\emph{"sub-coordinator"} $K_{2,3}$ of Figure~\ref{fig:figura1}).
$K_{2,3}$ represents the \emph{"unchangeable"}\footnote{Since we
want to be readily compose-able, our goal is to apply the
enhancements without modifying the coordinator and the components.}
environment that $K$ \emph{"offers"} to $W$. The wrapper $W$ is a
component whose interaction behavior is specified in each
enhancement of $E$. Depending on the logic it implements, we can
either built it by scratch or acquire it as a pre-existent $COTS$
component (e.g. a data translation component). $W$ intercepts the
messages exchanged between $K_{2,3}$, $C2$ and $C3$ and applies the
enhancements in $E$ on the interactions performed on the
communication channels 2 and 3 (i.e.: connectors 2 and 3 of
Figure~\ref{fig:figura1}). We first decouple $K$ (i.e.: $K_{2,3}$),
$C2$ and $C3$ to ensure that they no longer synchronize directly.
Then we automatically derive a behavioral model of $W$ (i.e.: a LTS)
from the bMSCs and HMSCs specification of $E$. We do this by
exploiting our implementation of the translation algorithm described
in~\cite{implied_scenarios}. Finally, if the insertion of $W$ in $S$
allows the resulting composed system (i.e.: $S^{\prime}$ after the
execution of the second step) to still satisfy each desired behavior
in $P$, $W$ is interposed between $K_{2,3}$, $C_2$ and $C_3$. To
insert $W$, we automatically synthesize two new coordinators
$K^\prime$ and $K^{\prime\prime}$. In general, $K^\prime$ always
refers to the coordinator between $W$ and the components affected by
the enhancement. $K^{\prime\prime}$ always refers to the coordinator
between $K$ and $W$. By referring to Section~\ref{CBA_architecture},
to do this, we automatically derive two behavioral models of $W$: i)
$W\_TOP$ which is the behavior of $W$ only related to its \emph{top}
interface and ii) $W\_BOTTOM$ which is the behavior of $W$ only
related to its \emph{bottom} interface.

In the second step, we derive the implementation of the synthesized
glue code used to insert $W$ in $S$. This glue code is the actual
code implementing $K^{\prime\prime}$ and $K^{\prime}$. By referring
to Figure~\ref{fig:figura1}, the parallel composition $K_{new}$ of
$K$, $K^{\prime}$, $K^{\prime\prime}$ and $W$ represents the
enhanced coordinator.\\

By iterating the whole approach, $K_{new}$ may be treated as $K$
with respect to the enforcing of new desired behaviors and the
application of new enhancements. This allows us to achieve
compose-ability of different coordinator protocol enhancements
(i.e.: modular protocol's transformations). In other words, our
approach is compositional in the automatic synthesis of the enhanced
glue code.

\section{Method formalization} \label{methodformalization}

In this section, by using an explanatory example, we formalize the
two steps of our method. For the sake of brevity we limit ourselves
to formalize the core of the extended approach. Refer
to~\cite{autili:tesi} for the formalization of the whole approach.

\begin{figure}
\centering \epsfig{file=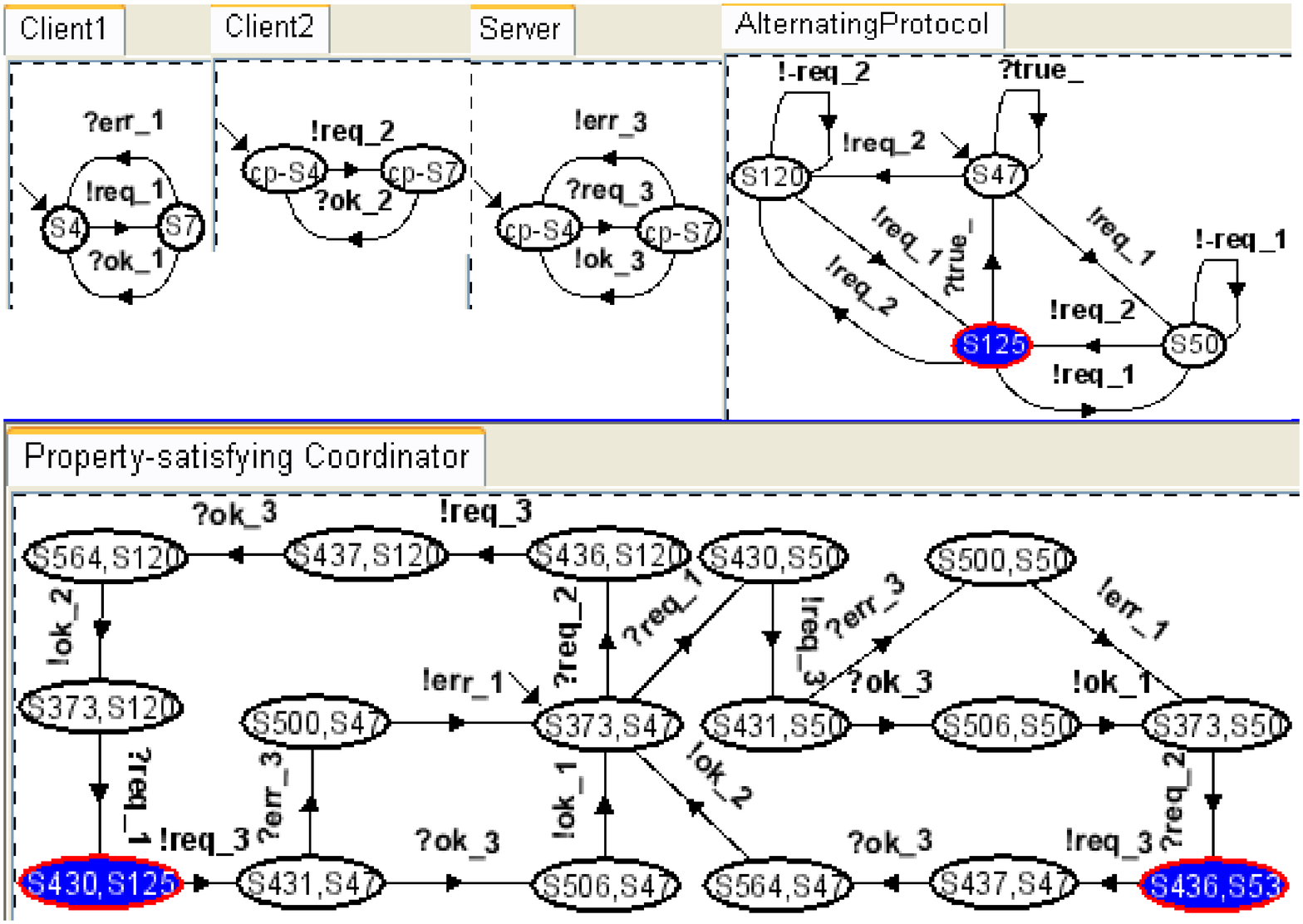, height=6cm,
width=9cm} \caption{$LTSs$ specification of $S$ and B\"{u}chi
Automata specification of $P$} \label{fig:oldsystem}
\end{figure}

In Figure~\ref{fig:oldsystem}, we consider screen-shots of the
\emph{SYNTHESIS} tool related to both the specification of $S$ and
of $P$. $Client1$, $Client2$ and $Server$ are the components in
$S$.\\\emph{Property-satisfying Coordinator} is the coordinator in
$S$ which satisfies the desired behavior denoted with
\emph{AlternatingProtocol}. In this example,
\emph{AlternatingProtocol} is the only element in the specification
$P$. The CBA configuration of $S$ is shown in the right-hand side of
Figure~\ref{cba_arch_sample} where $C_1$, $C_2$, $C_3$ and $K$ are
$Client1$, $Client2$, $Server$ and \emph{Property-satisfying
Coordinator} of Figure~\ref{fig:oldsystem} respectively.

Each LTS describes the behavior of a component or of a coordinator
instance in terms of the messages (seen as I/O actions) exchanged
with its environment\footnote{The environment of a
component/coordinator is the parallel composition of all others
components in the system.}. Each node is a state of the instance.
The node with the incoming arrow (e.g.: the state $S4$ of $Client1$
in Figure~\ref{fig:oldsystem}) is the starting state. An arc from a
node $n_1$ to a node $n_2$ denotes a transition from $n_1$ to $n_2$.
The transition labels prefixed by "!" denote output actions (i.e.:
sent requests and notifications), while the transition labels
prefixed by "?" denote input actions (i.e.: received requests and
notifications). In each transition label, the symbol "\_" followed
by a number denotes the identifier of the connector on which the
action has been performed. The filled nodes on the coordinator's LTS
denote states in which one execution of the behavior specified by
the B\"{u}chi Automaton \emph{AlternatingProtocol} has been
accomplished.

Each B\"{u}chi Automaton (see \emph{AlternatingProtocol} in
Figure~\ref{fig:oldsystem}) describes a desired behavior for $S$.
Each node is a state of $S$. The node with the incoming arrow is the
initial state. The filled nodes are the states accepting the desired
behavior. The syntax and semantics of the transition labels is the
same of the LTSs of components and coordinator except two kinds of
action: i) a universal action (e.g.: $?true\_$  in
Figure~\ref{fig:oldsystem}) which represents any possible
action\footnote{The prefixed symbols "!" or "?", in the label of a
universal action, are ignored by \emph{SYNTHESIS}.}, and ii) a
negative action (e.g.: $!-req\_2$ in Figure~\ref{fig:oldsystem})
which represents any possible action different from the negative
action itself\footnote{The prefixed symbols "!" or "?", in the label
of a negative action, are still interpreted by \emph{SYNTHESIS}.}.

$Client1$ performs a request (i.e.: action $!req\_1$) and waits for
a erroneous or successful notification: actions $?err\_1$ and
$?ok\_1$ respectively. $Client2$ simply performs the request and it
never handles erroneous notifications. $Server$ receives a request
and then it may answer either with a successful or an erroneous
notification~\footnote{The error could be either due to an
upper-bound on the number of request that $Server$ can accept
simultaneously or due to a general transient-fault on the
communication channel.}.

\emph{AlternatingProtocol} specifies the behavior of $S$ that
guarantees the evolution of all components. It specifies that
$Client1$ and $Client2$ must perform requests by using an
alternating coordination protocol. More precisely, if $Client1$
performs an action $req$ (the transition $!req\_1$ from the state
$S47$ to the state $S50$ in Figure~\ref{fig:oldsystem}) then it
cannot perform $req$ again (the loop transition $!-req\_1$ on the
state $S50$ in Figure~\ref{fig:oldsystem}) if $Client2$ has not
performed $req$ (the transition $!req\_2$ from the state $S50$ to
the accepting state $S125$ in Figure~\ref{fig:oldsystem}) and
viceversa.

\begin{figure}[h]
\centering \epsfig{file=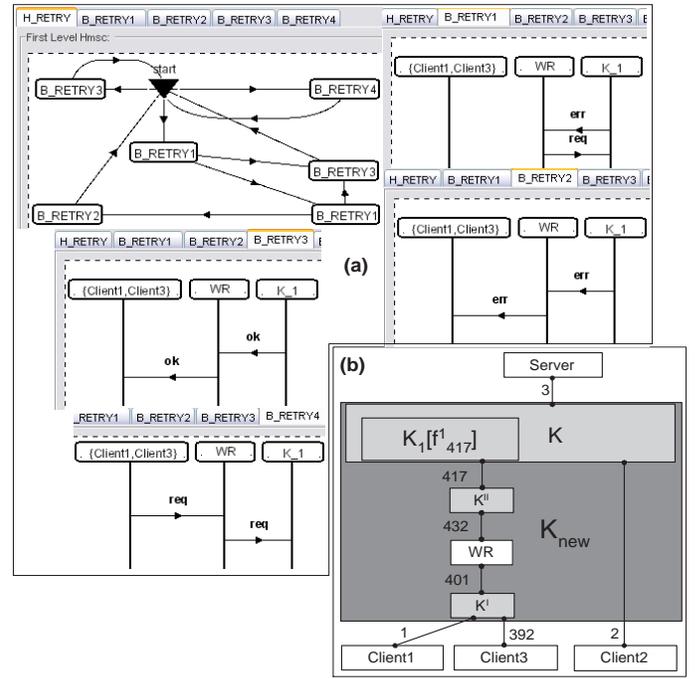, height=9cm, width=9cm}
\caption{$bMSCs$ and $hMSC$ specification of $E$: \emph{RETRY}
enhancement} \label{fig:retry}
\end{figure}

In Figure~\ref{fig:retry}.(a), we consider the specification of $E$
as given in input to the \emph{SYNTHESIS} tool. In this example, the
\emph{RETRY} enhancement is the only element in $E$.

$Client1$ is an interactive client and once an erroneous
notification occurs, it shows a dialog window displaying information
about the error. The user might not appreciate this error message
and he might lose the degree of trust in the system. By recalling
that the dependability of a system reflects the user's degree of
trust in the system, this example shows a commonly practiced
dependability-enhancing technique. The wrapper \emph{WR} attempts to
hide the error to the user by re-sending the request a finite number
of times. This is the \emph{RETRY} enhancement specified in
Figure~\ref{fig:retry}.(a). The wrapper \emph{WR} re-sends at most
two times. Moreover, the \emph{RETRY} enhancement specifies an
update of $S$ obtained by inserting $Client3$ which is a new client.
In specifying enhancements, we use \emph{"augmented"}-bMSCs. By
referring to~\cite{itu:msc}, each usual bMSC represents a possible
execution scenario of the system. Each execution scenario is
described in terms of a set of interacting components, sequences of
method call and possible corresponding return values. To each
vertical lines is associated an instance of a component. Each
horizontal arrow represents a method call or a return value. Each
usual HMSC describes possible continuations from a scenario to
another one. It is a graph with two special nodes: the starting and
the ending node. Each other node is related to a specified scenario.
An arrow represents a transition from a scenario to another one. In
other words, each HMSC composes the possible execution scenarios of
the system. The only difference between \emph{"augmented"}-bMSCs and
usual bMSCs is that to each vertical line can be associated a set of
component instances (e.g.: $\{Client1, Client3\}$ in
Figure~\ref{fig:retry}.(a)) rather than only one instance. This is
helpful when we need to group components having the same interaction
behavior.

\subsection{First step: wrapper insertion procedure} \label{first_step}

By referring to Section~\ref{methoddescription}, each enhancement
MSCs specification (see Figure~\ref{fig:retry}.(a)) is in general
described in terms of the \emph{sub-coordinator} $K_1$ (i.e.: $K\_1$
in Figure~\ref{fig:retry}.(a)), the wrapper (\emph{WR}), the
components in $S$ ($Client1$) and the new components ($Client3$).
The LTS of the \emph{sub-coordinator} is automatically derived from
the LTS of the coordinator in $S$ ($K$) by performing the following
algorithm:

\begin{definition} \label{k_restriction}
\emph{$L_{j,..,j+h}$ construction algorithm}
\\Let $L$ be the
LTS for a component (or a coordinator) $C$, we derive the LTS
$L_{j,..,j+h}$, $h\geq0$, of the \emph{behavior of $C$ on channels
j,..,j+h} as follows:

\begin{enumerate}
\item set $L_{j,..,j+h}$ equal to $L$;

\item for each loop $(\nu,\nu)$ of $L_{j,..,j+h}$ labeled with an action $\alpha = a_k$ where $k\neq
j,..,j+h$ do:\\
$~~~~$remove $(\nu,\nu)$ from the set of arcs of $L_{j,..,j+h}$;

\item for each arc $(\nu,\mu)$ of $L_{j,..,j+h}$ labeled with an action $\alpha = a_k$ where $k\neq
j,..,j+h$ do:
\begin{itemize}
\item remove $(\nu,\mu)$ from the set of arcs of $L_{j,..,j+h}$;
\item if $\mu$ is the starting state then\\
$~~~~$set $\nu$ as the starting state;
\item for each other arc $(\nu,\mu)$ of $L_{j,..,j+h}$ do:\\
$~~~~$ replace $(\nu,\mu)$ with $(\nu,\nu)$;
\item for each arc $(\mu,\nu)$ of $L_{j,..,j+h}$ do:\\
$~~~~$ replace $(\mu,\nu)$ with $(\nu,\nu)$;
\item for each arc $(\mu,\upsilon)$ of $L_{j,..,j+h}$ with $\upsilon\neq\mu,\nu$ do:\\
$~~~~$ replace $(\mu,\upsilon)$ with $(\nu,\upsilon)$;
\item for each arc $(\upsilon,\mu)$ of $L_{j,..,j+h}$ with $\upsilon\neq\mu,\nu$ do:\\
$~~~~$ replace $(\upsilon,\mu)$ with $(\upsilon,\nu)$;
\item for each loop $(\mu,\mu)$ of $L_{j,..,j+h}$ do:\\
$~~~~$ replace $(\mu,\mu)$ with $(\nu,\nu)$;
\item remove $\mu$ from the set of nodes of $L_{j,..,j+h}$;
\end{itemize}

\item until $L_{j,..,j+h}$ is a non-deterministic LTS (i.e.: it contains arcs labeled with the same action and outgoing the same
node) do:
\begin{itemize}
\item for each pair of loops $(\nu,\nu)$ and $(\nu,\nu)$ of $L_{j,..,j+h}$ labeled with the same action
do:\\
$~~~~$ remove $(\nu,\nu)$ from the set of arcs of $L_{j,..,j+h}$;

\item for each pair of arcs $(\nu,\mu)$ and $(\nu,\mu)$ of $L_{j,..,j+h}$ labeled with the same action
do:\\
$~~~~$ remove $(\nu,\mu)$ from the set of arcs of $L_{j,..,j+h}$;

\item for each pair of arcs ($(\nu,\mu)$ and $(\nu,\upsilon)$) or ($(\nu,\nu)$ and $(\nu,\upsilon)$) of $L_{j,..,j+h}$ labeled with the same action do:
\begin{itemize}
\item remove $(\nu,\upsilon)$ from the set of arcs of $L_{j,..,j+h}$;
\item if $\upsilon$ is the starting state then\\
$~~~~$set $\nu$ as the starting state;
\item for each ingoing arc $in$ in $\upsilon$, outgoing arc $out$
from $\upsilon$ and loop $l$ on $\upsilon$ do:\\
$~~~~$move the extremity on $\upsilon$ of $in$, $out$ and $l$\\
$~~~~$on $\nu$;
\item remove $\upsilon$ from the set of nodes of $L_{j,..,j+h}$.
\end{itemize}
\end{itemize}
\end{enumerate}
\end{definition}

\noindent Informally, the algorithm of
Definition~\ref{k_restriction} \emph{"collapses"} (steps 1,2 and 3)
linear and/or cyclic paths made only of actions on channels $k$
$\neq$ $j,..,j+h$. Moreover, it also avoids (step 4) possible
\emph{"redundant"} non-deterministic behaviors\footnote{These
behaviors might be a side effect due to the collapsing.}.

\begin{figure}
\centering \epsfig{file=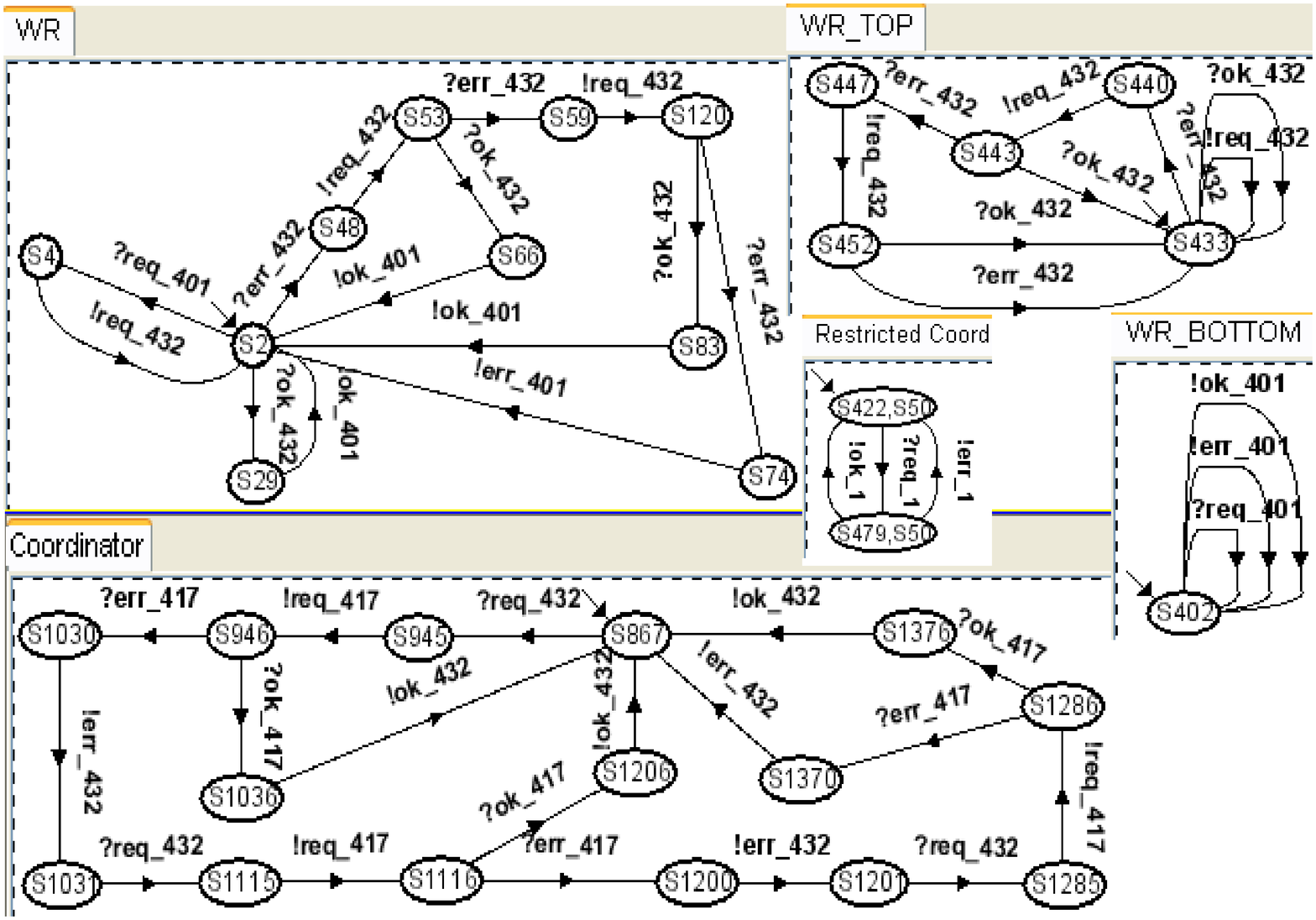, height=6.5cm,
width=9cm} \caption{LTSs of wrapper, sub-coordinator and
$K^{\prime\prime}$} \label{fig:newsystem}
\end{figure}

By referring to Figure~\ref{fig:newsystem}, the LTS of $K_1$ is the
$LTS$ \emph{Restricted Coord}.

In general, once we derived $K_{j,..,j+h}$, we decouple $K$ from the
components $C_j,..,C_{j+h}$ (i.e.: $Client1$) connected through the
connectors $j,..,j+h$ which are related to the specified enhancement
(i.e.: the connector 1). To do this, we use the decoupling function
defined as follows:

\begin{definition} \label{k_decoupling} \emph{Decoupling function}\\
Let $Act_{K}$ be the set of action labels of the coordinator $K$,
let be $ID$ $=$ $\{j,\ldots,j+h\}$ a subset of all connectors
identifiers\footnote{By referring to
Section~\ref{methodformalization}, the connectors identifiers are
postfixed to the labels in $Act_{K}$.} of $K$ and let be $\delta$
$\neq$ $j,..,j+h$ a new connector identifier, we define the
\emph{"decoupling"} function $f^{j,..,j+h}_{\delta}$ as follows:
\begin{itemize}
\item $\forall$ $a_i$ $\in$ $Act_{K}$, if $i \in ID$ then: $f^{j,..,j+h}_{\delta}(a_i)$ $=$ $a_{\delta};$
\end{itemize}
\end{definition}

\noindent The unique connector identifier $\delta$ is automatically
generated by \emph{SYNTHESIS}. In this way we ensure that $K$ and
$Client1$ no longer synchronize directly. In~\cite{companion}, we
detail the correspondence between the decoupling function and
components/coordinator deployment.

\noindent Now, by continuing the method described in
Section~\ref{methoddescription}, we derive the LTSs for the wrapper
(see \emph{WR} in Figure~\ref{fig:newsystem}) and the new components
(the $LTS$ of $Client3$ is equal to the LTS of $Client1$ in
Figure~\ref{fig:oldsystem} except for the connector identifier). We
recall that \emph{SYNTHESIS} does that by taking into account the
enhancements specification and by performing its implementation of
the translation algorithm described in~\cite{implied_scenarios}. It
is worthwhile noticing that \emph{SYNTHESIS} automatically generates
the connector identifiers for the actions performed by \emph{WR} and
$Client3$. By referring to Section~\ref{CBA_architecture}, \emph{WR}
is connected to its environment through two connectors: i) one on
its top interface (i.e.: the connector $432$ in
Figure~\ref{fig:retry}.(b)) and ii) one on its bottom interface
(i.e.: the connector $401$ in Figure~\ref{fig:retry}.(b)). Finally,
as we will see in detail in Section~\ref{check}, if the insertion of
\emph{WR} allows the resulting composed system to still satisfy
\emph{AlternatingProtocol}, \emph{WR} is interposed between
$K_{1}[f^{1}_{417}]$, $Client1$ and $Client3$. We recall that
$K_{1}[f^{1}_{417}]$ is $K_{1}$ (see \emph{Restricted Coord} in
Figure~\ref{fig:newsystem}) renamed after the decoupling. To insert
\emph{WR}, \emph{SYNTHESIS} automatically synthesizes two new
coordinators $K^\prime$ and $K^{\prime\prime}$. \emph{Coordinator}
in Figure~\ref{fig:newsystem} is the LTS for $K^{\prime\prime}$. For
the purposes of this paper, in Figure~\ref{fig:newsystem}, we omit
the LTS for $K^{\prime}$. $K^{\prime\prime}$ is derived by taking
into account both the LTSs of $K_{1}[f^{1}_{417}]$ and
\emph{WR\_TOP} (see Figure~\ref{fig:newsystem}) and by performing
the old synthesis approach~\cite{bertinoro,cbse7,savcbs03}. While
$K^\prime$ is derived analogously to the LTSs of $Client1$,
$Client3$ and \emph{WR\_BOTTOM} (see Figure~\ref{fig:newsystem}).
The LTSs of \emph{WR\_TOP} and \emph{WR\_BOTTOM} are
\emph{WR$_{432}$} and \emph{WR$_{401}$} respectively. By referring
to Definition~\ref{k_restriction}, \emph{WR$_{432}$} and
\emph{WR$_{401}$} model the behavior of \emph{WR} on channels 432
and 401 respectively. The resulting enhanced, deadlock-free and
\emph{Alternating Protocol}-satisfying system is\\ $S^{\prime}$
$\equiv$ $((Client_1 \mid$ $Client_2 \mid$ $Client_3 \mid$ $Server
\mid$ $K_{new})$ $\backslash$ $(Act_{Client1} \cup$ $Act_{Client2}
\cup$ $Act_{Server} \cup$ $Act_{Client3}))$ where $K_{new}$ $\equiv$
$((K[f^{1}_{417}] \mid$ $WR \mid$ $K^{\prime} \mid$
$K^{\prime\prime})$ $\backslash$ $(Act_{WR} \cup$
$Act_{K_1[f^{1}_{417}]}))$ (see Figure~\ref{fig:retry}.(b)).
In~\cite{autili:tesi}, we proved the correctness and completeness of
the approach.

\subsection{Second step: synthesis of the glue code implementation} \label{second_step}

The parallel composition $K_{new}$ represents the model of the
enhanced coordinator. By referring to
Section~\ref{methoddescription}, we recall that \emph{K} is the
initial glue code for $S$ and \emph{WR} is a COTS component whose
interaction behavior is specified by the enhancements specification
$E$. That is, the actual code for \emph{WR} and $K$ is already
available. Thus, in order to derive the code implementing $K_{new}$,
\emph{SYNTHESIS} automatically derives the actual code implementing
$K^{\prime}$ and $K^{\prime\prime}$. Using the same technique
described in~\cite{bertinoro,cbse7,savcbs03,companion}, this is done
by exploiting the information stored in the nodes and arcs of the
LTSs for $K^{\prime}$ and $K^{\prime\prime}$. More precisely, the
code implementing $K^{\prime}$ and $K^{\prime\prime}$ reflects the
structure of their LTSs which describe state machines. For the sake
of brevity, here, we omit a detailed description of the code
synthesis. Refer to~\cite{bertinoro,cbse7,savcbs03,companion} for
it. In~\cite{cbse7,companion}, we validated and applied
\emph{SYNTHESIS} for assembling \emph{Microsoft COM/DCOM}
components. The reference development platform of the current
version of \emph{SYNTHESIS} is \emph{Microsoft Visual Studio 7.0}
with \emph{Active Template Library}.

\section{Checking enhancement\\ consistency} \label{check}

In this section we formalize a compositional technique to check if
the applied enhancements are consistent with respect to the
previously enforced desired behaviors. In other words, given the
B\"{u}chi Automata specification of a desired behavior $P_i$, given
the deadlock-free and $P_i$-satisfying coordinator $K$ and given the
MSCs specification of an enhancement $E_i$, we check (in a
compositional way) if the enhanced coordinator $K_{new}$ still
satisfies $P_i$ ($K_{new}$ $\models$ $P_i$).

In general, we have to check $((K[f^{j,..,j+h}_{\delta}] \mid$ $WR
\mid$ $K^{\prime} \mid$ $K^{\prime\prime})$ $\backslash$ $(Act_{WR}
\cup$ $Act_{K_{j,..,j+h}[f^{j,..,j+h}_{\delta}]}))$ $\models$ $P_i$.
By exploiting the constraints of our architectural style, it is
enough to check $((K[f^{j,..,j+m}_{\delta}] \mid$
$K^{\prime\prime}_{\delta})$ $\backslash$
$Act_{K_{j,..,j+m}[f^{j,..,j+m}_{\delta}]})$ $\models$ $P_i$ where
$\{j,..,j+m\}$ is the set of channel identifiers which are both
channels in $\{j,..,j+h\}$ and in the set of channel identifiers for
the action labels in $P_i$. In order to avoid the state explosion
phenomenon we should decompose the verification without composing in
parallel the processes $K[f^{j,..,j+m}_{\delta}]$ and
$K^{\prime\prime}_{\delta}$. We do that by exploiting the
\emph{assume-guarantee paradigm} for compositional
reasoning~\cite{Clarke99}.

By recasting the typical proof strategy of the
\emph{assume-guarantee paradigm} in our context, we know that if\\
$\langle A \rangle K[f^{j,..,j+m}_{\delta}] \langle P_i \rangle$ and
$\langle true \rangle K^{\prime\prime}_{\delta} \langle A \rangle$
hold then we can conclude that\\ $\langle true \rangle$
$((K[f^{j,..,j+m}_{\delta}] \mid$ $K^{\prime\prime}_{\delta})$
$\backslash$ $Act_{K_{j,..,j+m}[f^{j,..,j+m}_{\delta}]})$ $\langle
P_i \rangle$ is true. This proof strategy can also be expressed as
the following inference rule:\\

\begin{figure}[h]
\centering \epsfig{file=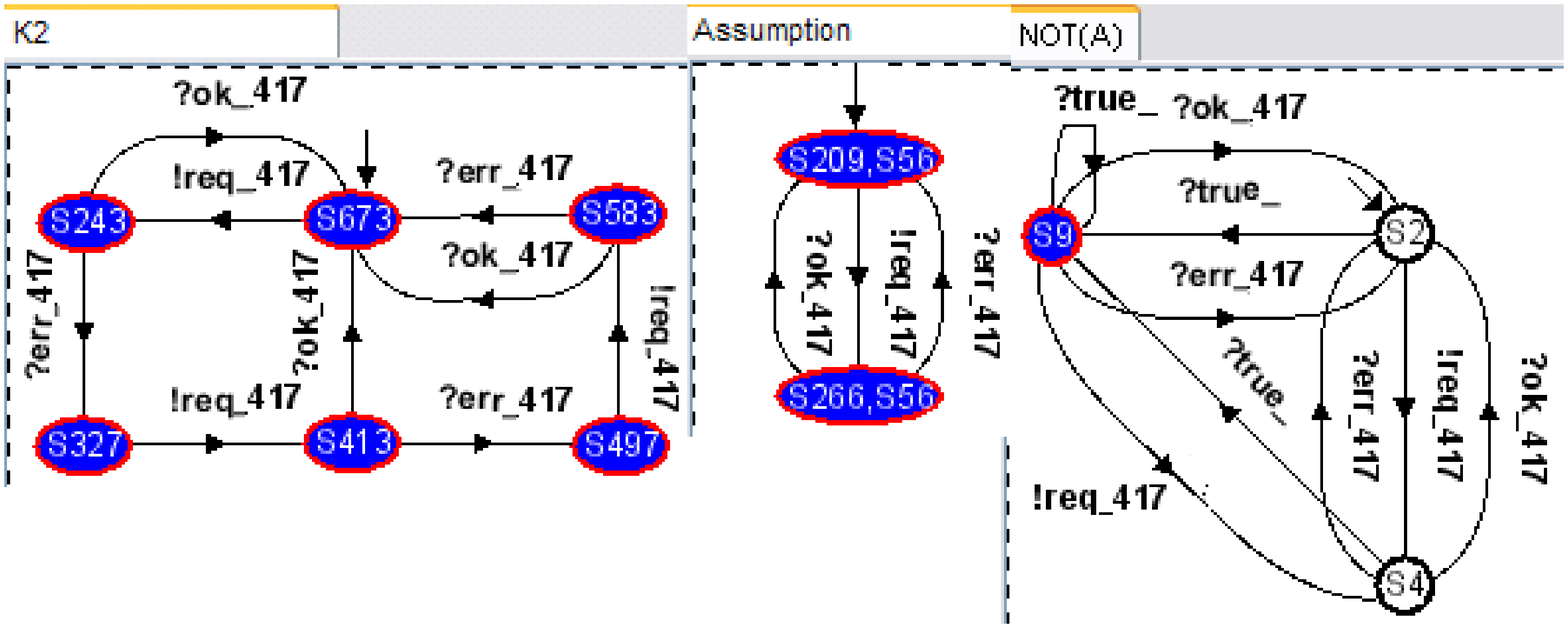, height=4cm,
width=8.5cm} \caption{B\"{u}chi Automata of
$K^{\prime\prime}_{417}$, $K_{1}[f^{1}_{417}][f_{env}]$ and
$\overline{K_{1}[f^{1}_{417}][f_{env}]}$} \label{A_and_K2}
\end{figure}

\begin{center}
\begin{small}
$\langle true \rangle K^{\prime\prime}_{\delta} \langle A \rangle$\\
\end{small}
\begin{large}
$\frac{\langle A \rangle K[f^{j,..,j+m}_{\delta}] \langle P_i
\rangle}{\langle true \rangle((K[f^{j,..,j+m}_{\delta}] \mid
K^{\prime\prime}_{\delta})\backslash
Act_{K_{j,..,j+m}[f^{j,..,j+m}_{\delta}]})\langle P_i \rangle}$
\end{large}
\end{center}

\noindent where $A$ is a LTL formula (and hence it is modeled as a
B\"{u}chi Automaton). We recall that, in $S$, $K$ already satisfies
$P_i$. Once we applied $E_i$ to obtain the enhanced system
$S^{\prime}$, $A$ represents the assumptions (in $S^{\prime}$) on
the environment of $K$ that must be held on the channel $\delta$ in
order to make $K$ able to still satisfy $P_i$. Without loss of
generality, let $\{j,..,j+m\}$ be $Channels_{P_i}$ $\cap$
$\{j,..,j+h\}$ where $Channels_{P_i}$ is the set of channel
identifiers for the action labels in $P_i$; then $A$ is the
B\"{u}chi Automaton corresponding to
$K_{j,..,j+m}[f^{j,..,j+m}_{\delta}][f_{env}]$ where
$f_{env}(?\alpha)$ $=$ $!\alpha$ and $f_{env}(!\alpha)$ $=$
$?\alpha$. For the example illustrated in
Section~\ref{methodformalization}, $K^{\prime\prime}_{\delta}$ and
$A$ are the B\"{u}chi Automata corresponding to
$K^{\prime\prime}_{417}$ (i.e.: $K2$ showed in
Figure~\ref{A_and_K2}) and to $K_{1}[f^{1}_{417}][f_{env}]$
($Assumption$ showed in Figure~\ref{A_and_K2}) respectively.

\noindent In general, a formula $\langle true \rangle M \langle P
\rangle$ means $M \models P$. While a formula $\langle A \rangle M
\langle P \rangle$ means if $A$ holds then $M \models P$. In our
context, $P$ is modeled as the corresponding B\"{u}chi Automaton
$B_{P}$. $M$ is modeled as the corresponding LTS. By referring
to~\cite{Clarke99}, to a LTS $M$ always corresponds a B\"{u}chi
Automaton $B_{M}$. With $L(B)$ we denote the language accepted by
$B$. Exploiting the usual automata-based model checking
approach~\cite{Clarke99}, to check if $M \models P$ we first
automatically build the product language $L_{M \cap P}$ $\equiv$
$L(B_{M})$ $\cap$ $\overline{L(B_{P})}$ and then we check if $L_{M
\cap P}$ is empty.

\begin{theorem} \label{assume} \emph{Enhancement consistency check}\\
Let $P_i$ be the B\"{u}chi Automata specification of a desired
behavior for a system $S$ formed by $C_1,..,C_n$ components; let $K$
be the deadlock-free and $P_i$-satisfying coordinator for the
components in $S$; let $E_i$ be the MSCs specification of a
$K$-protocol enhancement; let $K^{\prime\prime}$ be the adaptor
between $K$ and the wrapper implementing the enhancement $E_i$; let
$\delta$ the identifier of the channel connecting $K^{\prime\prime}$
with $K$; let $\{j,..,j+m\}$ the set of channel identifiers which
are both channels in the set of channel identifiers for the action
labels in $P_i$ and in the set of channels identifiers affected by
the enhancement $E_i$; and let $f_{env}$ be a relabeling function in
such a way that $f_{env}(?\alpha)$ $=$ $!\alpha$ and
$f_{env}(!\alpha)$ $=$ $?\alpha$ for all $\alpha\in Act_{K}$, if
$L_{K^{\prime\prime}_{\delta} \cap
K_{j,..,j+m}[f^{j,..,j+m}_{\delta}][f_{env}]} = \emptyset$ then\\
$((K[f^{j,..,j+m}_{\delta}] \mid$ $K^{\prime\prime}_{\delta})$
$\backslash$ $Act_{K_{j,..,j+m}[f^{j,..,j+m}_{\delta}]})$ $\models$
$P_i$ and hence $E_i$ is consistent with respect to $P_i$.
\begin{proof}
Let $A$ be the B\"{u}chi Automaton corresponding to
$K_{j,..,j+m}[f^{j,..,j+m}_{\delta}][f_{env}]$, if
$L_{K^{\prime\prime}_{\delta} \cap
K_{j,..,j+m}[f^{j,..,j+m}_{\delta}][f_{env}]} = \emptyset$ then
$K^{\prime\prime}_{\delta}$ $\models$ $A$. That is, $\langle true
\rangle K^{\prime\prime}_{\delta} \langle A \rangle$ holds.
Moreover, by construction of $A$, $\langle A \rangle
K[f^{j,..,j+m}_{\delta}] \langle P_i \rangle$ holds too. By applying
the inference rule of the \emph{assume-guarantee paradigm}, $\langle
true \rangle((K[f^{j,..,j+m}_{\delta}] \mid
K^{\prime\prime}_{\delta})\backslash
Act_{K_{j,..,j+m}[f^{j,..,j+m}_{\delta}]})\langle P_i \rangle$ is
true and hence $((K[f^{j,..,j+m}_{\delta}] \mid$
$K^{\prime\prime}_{\delta})$ $\backslash$
$Act_{K_{j,..,j+m}[f^{j,..,j+m}_{\delta}]})$ $\models$ $P_i$.
\end{proof}
\end{theorem}

\noindent By referring to Theorem~\ref{assume}, to check if $E_i$ is
consistent with respect to $P_i$, it is enough to check if $\langle
true \rangle K^{\prime\prime}_{\delta} \langle A \rangle$ holds. In
other words, it is enough to check if $K^{\prime\prime}$ provides
$K$ with the environment it expects (to still satisfy $P_i$) on the
channel connecting $K^{\prime\prime}$ to $K$ (i.e.: the connector
identified by $\delta$). In the example illustrated in
Section~\ref{methodformalization}, \emph{RETRY} is consistent with
respect to \emph{AlternatingProtocol}. In fact, by referring to
Figure~\ref{A_and_K2}, $NOT(A)$ is the B\"{u}chi Automaton for
$\overline{K_{1}[f^{1}_{417}][f_{env}]}$ (i.e.: for $\overline{A}$
of Theorem~\ref{assume}) and $K2$ is the B\"{u}chi Automaton for
$K^{\prime\prime}_{417}$ (i.e.: for $K^{\prime\prime}_{\delta}$ of
Theorem~\ref{assume}). By automatically building the product
language between the languages accepted by $K2$ and $NOT(A)$,
\emph{SYNTHESIS} concludes that $L_{K^{\prime\prime}_{417} \cap
K_{1}[f^{1}_{417}][f_{env}]} = \emptyset$ and hence that\\
$((K[f^{1}_{417}] \mid$ $K^{\prime\prime}_{417})$ $\backslash$
$Act_{K_{1}[f^{1}_{417}]})$ $\models$ \emph{AlternatingProtocol}.
That is \emph{RETRY} is consistent with respect to
\emph{AlternatingProtocol}.

\section{Conclusion and future work} \label{conclusion}

In this paper, we combined the approaches of protocol transformation
formalization~\cite{wrappers_formalization} and of automatic
coordinator synthesis~\cite{bertinoro,cbse7,savcbs03} to produce a
new technique for automatically synthesizing failure-free
coordinators for protocol enhanced in component-based systems. The
two approaches take advantage of each other: while the approach of
protocol transformations formalization adds compose-ability to the
automatic coordinator synthesis approach, the latter adds automation
to the former. This paper is a revisited and extended version
of~\cite{wcat04}. With respect to~\cite{wcat04}, the novel aspects
of this work are that we have definitively fixed and extended the
formalization of the approach, we have implemented it in our
\emph{"SYNTHESIS"} tool and we have formalized and implemented the
enhancement consistency check.

The key results are: (i) the extended approach is compositional in
the automatic synthesis of the enhanced coordinator; that is, each
wrapper represents a modular protocol transformation so that we can
apply coordinator protocol enhancements in an incremental way by
re-using the code synthesized for already applied enhancements; (ii)
we are able to add extra functionality to a coordinator beyond
simply restricting its behavior;(iii) this, in turn, allows us to
enhance a coordinator with respect to a useful set of protocol
transformations such as the set of transformations referred
in~\cite{wrappers_formalization}. The automation and applicability
of both the old (presented in~\cite{bertinoro,cbse7,savcbs03}) and
the extended (presented in this paper and in~\cite{companion})
approach for synthesizing coordinators is supported by our tool
called \emph{"SYNTHESIS"}~\cite{cbse7,companion}.

As future work, we plan to: (i) develop more user-friendly
specification of both the desired behaviors and the protocol
enhancements (e.g., UML2 Interaction Overview Diagrams and Sequence
Diagrams); (ii) validate the applicability of the whole approach to
large-scale examples different than the case-study treated
in~\cite{companion} which represents the first attempt to apply the
extended version of \emph{"SYNTHESIS"} (formalized in this paper) in
real-scale contexts.

\bibliographystyle{abbrv}
\bibliography{savcbs04}

\begin{thebibliography}{10}

\bibitem{autili:tesi}
M.~Autili.
\newblock Sintesi automatica di connettori per protocolli di comunicazione
  evoluti.
\newblock Tesi di laurea in Informatica, Universit\'a dell'Aquila - April,2004
  - http://www.di.univaq.it/tivoli/AutiliThesis.pdf.

\bibitem{supervisory1}
S.~Balemi, G.~J. Hoffmann, P.~Gyugyi, H.~Wong-Toi, and G.~F. Franklin.
\newblock Supervisory control of a rapid thermal multiprocessor.
\newblock {\em IEEE Transactions on Automatic Control}, 38(7):1040--1059, July
  1993.

\bibitem{interface_automata}
L.~de~Alfaro and T.~Heinzinger.
\newblock Interface automata.
\newblock In {\em ACM Proc. of the joint 8th ESEC and 9th FSE}, 2001.

\bibitem{bertinoro}
P.~Inverardi and M.~Tivoli.
\newblock {\em Software Architecture for Correct Components Assembly - Chapter
  in: Formal Methods for the Design of Computer, Communication and Software
  Systems: Software Architecture}.
\newblock Springer, LNCS 2804, Sept. 2003.

\bibitem{itu:msc}
ITU-T.
\newblock Reccomendation z.120. message sequence charts. (msc'96), 1996.

\bibitem{Clarke99}
E.~M.~C. Jr., O.~Grumberg, and D.~A. Peled.
\newblock {\em Model Checking}.
\newblock The MIT Press, 2001.

\bibitem{wcat04}
M.Autili, P.Inverardi, and M.Tivoli.
\newblock Automatic adaptor synthesis for protocol transformation.
\newblock In {\em WCAT04}, 2004.

\bibitem{milner:ccs}
R.~Milner.
\newblock {\em Communication and Concurrency}.
\newblock Prentice Hall, New York, 1989.

\bibitem{cbse7}
M.Tivoli, P.Inverardi, V.Presutti, A.Forghieri, and M.Sebastianis.
\newblock Correct components assembly for a product data management cooperative
  system.
\newblock In {\em proceedings of the Int. Symposium CBSE7. May,2004}. Springer,
  LNCS 3054, 2007.

\bibitem{converter_synthesis}
R.~Passerone, L.~de~Alfaro, T.~Heinzinger, and A.~L. Sangiovanni-Vincentelli.
\newblock Convertibility verification and converter synthesis: Two faces of the
  same coin.
\newblock In {\em Proc. of ICCAD}, 2002.

\bibitem{savcbs03}
P.Inverardi and M.Tivoli.
\newblock Failure-free connector synthesis for correct components assembly.
\newblock In {\em Proceedings of SAVCBS'03}, 2003.

\bibitem{GS96}
M.~Shaw and D.~Garlan.
\newblock {\em Software Architecture: Perspectives on an Emerging Discipline}.
\newblock Prentice Hall, 1996.

\bibitem{wrappers_formalization}
B.~Spitznagel and D.~Garlan.
\newblock A compositional formalization of connector wrappers.
\newblock In {\em proceeding of the 25th ICSE'03 - Portland, OG (USA)}, May
  2003.

\bibitem{Szy98}
C.~Szyperski.
\newblock {\em Component Software. Beyond Object Oriented Programming}.
\newblock Addison Wesley, 1998.

\bibitem{companion}
M.~Tivoli, M.~Autili, and P.~Inverardi.
\newblock Synthesis: a tool for synthesizing correct and protocol-enhanced
  adaptors.
\newblock submitted for publication - Aug,2004 -
  http://www.di.univaq.it/tivoli/LastSynthesis.pdf.

\bibitem{CMU_techrep}
M.~Tivoli and D.~Garlan.
\newblock Coordinator synthesis for reliability enhancement in component-based
  systems.
\newblock Carnegie Mellon University, C.S.Dep. - Tech.Rep. -
  http://www.di.univaq.it/tivoli/CMUtechrep.pdf.

\bibitem{implied_scenarios}
S.~Uchitel, J.~Kramer, and J.~Magee.
\newblock Detecting implied scenarios in message sequence chart specifications.
\newblock In {\em ACM Proceedings of the joint 8th ESEC and 9th FSE}, Vienna,
  Sep 2001.

\bibitem{Yel97}
D.~M. Yellin and R.~E. Strom.
\newblock Protocol specifications and component adaptors.
\newblock {\em ACM Transactions on Programming Languages and Systems},
  19(2):292--333, march 1997.

\end{thebibliography}

\end{document}